\documentclass[12pt]{article}
\usepackage{graphicx,amssymb,amsfonts,amsmath,epsfig,color}
\usepackage{mathrsfs}
\usepackage{cite}
\usepackage{url}
\usepackage{float}
\newcounter{mnotecount}

\newcommand{\mnotex}[1]
{\protect{\stepcounter{mnotecount}}$^{\mbox{\footnotesize $\bullet$\themnotecount}}$
\marginpar{
\raggedright\tiny\em
$\!\!\!\!\!\!\,\bullet$\themnotecount: #1} }

\makeatletter
\DeclareSymbolFont{AMSb}{U}{msb}{m}{n}
\DeclareSymbolFontAlphabet{\mathbb}{AMSb}
\renewcommand{\section}{\@startsection{section}{1}{\z@}%
                                    {-7ex \@plus -1ex \@minus -.2ex}%
                                    {2.5ex \@plus.2ex}%
                                    {\normalfont\large\scshape\centering}}
\renewcommand{\subsection}{\@startsection{subsection}{2}{\z@}%
                                       {-5ex \@plus -1ex \@minus -.2ex}%
                                       {1.5ex \@plus.2ex}%
                                       {\normalfont\normalsize\scshape}}
\renewcommand{\subsubsection}{\@startsection{subsubsection}{3}{\z@}%
                                       {-5ex \@plus -1ex \@minus -.2ex}%
                                       {1.5ex \@plus.2ex}%
                                       {\normalfont\normalsize\scshape}}

\renewcommand\@seccntformat[1]{\ignorespaces\csname #1name\endcsname\space
                               \csname the#1\endcsname.\quad}   
\newdimen\captionmargin
\newdimen\captionindent
\newdimen\captionwidth
\newcommand{\captionfont}{\slshape}
\newcommand\@captionlabel[1]{\textsc{#1:}\space}
\long\def\@makecaption#1#2{%
  \vskip\abovecaptionskip
  \captionwidth\hsize
  \advance\captionwidth -2\captionmargin
  \sbox\@tempboxa{\@captionlabel{#1}\captionfont #2}%
  \ifdim \wd\@tempboxa >\captionwidth
    \ifdim\captionindent>\z@
      \advance\captionwidth -\captionindent
      \hskip\captionindent
    \fi
    \hskip\captionmargin
    \parbox[t]{\captionwidth}{\leavevmode\hskip-\captionindent
      \@captionlabel{#1}\captionfont #2}%
  \else
    \global \@minipagefalse
    \hb@xt@\hsize{\hfil\box\@tempboxa\hfil}%
  \fi
  \vskip\belowcaptionskip}
\def\eqnarray{%
   \stepcounter{equation}%
   \def\@currentlabel{\p@equation\theequation}%
   \global\@eqnswtrue
   \m@th
   \global\@eqcnt\z@
   \tabskip\@centering
   \let\\\@eqncr
   $$\everycr{}\halign to\displaywidth\bgroup
       \hskip\@centering$\displaystyle\tabskip\z@skip{##}$\@eqnsel
      &\global\@eqcnt\@ne$\;\hfil{##}$\hfil
      &\global\@eqcnt\tw@$\;\displaystyle{##}$\hfil\tabskip\@centering
      &\global\@eqcnt\thr@@ \hb@xt@\z@\bgroup\hss##\egroup
         \tabskip\z@skip
      \cr}
\ifcase \@ptsize
\or
\or
\fi
  \divide\@tempdima\baselineskip
  \@tempcnta=\@tempdima
\makeatother
\begin{document}

\renewcommand{\theequation}{\arabic{section}.\arabic{equation}}
\renewcommand{\thefigure}{\arabic{figure}}
\newcommand{\gapprox}{%
\mathrel{%
\setbox0=\hbox{$>$}\raise0.6ex\copy0\kern-\wd0\lower0.65ex\hbox{$\sim$}}}
\textwidth 165mm \textheight 220mm \topmargin 0pt \oddsidemargin 2mm
\def\ib{{\bar \imath}}
\def\jb{{\bar \jmath}}

\newcommand{\ft}[2]{{\textstyle\frac{#1}{#2}}}
\newcommand{\be}{\begin{equation}}
\newcommand{\ee}{\end{equation}}
\newcommand{\bea}{\begin{eqnarray}}
\newcommand{\eea}{\end{eqnarray}}
\newcommand{\Identity}{{1\!\rm l}}
\newcommand{\cx}{\overset{\circ}{x}_2}
\def\CN{$\mathcal{N}$}
\def\CH{$\mathcal{H}$}
\def\hg{\hat{g}}
\newcommand{\bref}[1]{(\ref{#1})}
\def\espai{\;\;\;\;\;\;}
\def\zespai{\;\;\;\;}
\def\avall{\vspace{0.5cm}}
\newtheorem{theorem}{Theorem}
\newtheorem{acknowledgement}{Acknowledgment}
\newtheorem{algorithm}{Algorithm}
\newtheorem{axiom}{Axiom}
\newtheorem{case}{Case}
\newtheorem{claim}{Claim}
\newtheorem{conclusion}{Conclusion}
\newtheorem{condition}{Condition}
\newtheorem{conjecture}{Conjecture}
\newtheorem{corollary}{Corollary}
\newtheorem{criterion}{Criterion}
\newtheorem{defi}{Definition}
\newtheorem{example}{Example}
\newtheorem{exercise}{Exercise}
\newtheorem{lemma}{Lemma}
\newtheorem{notation}{Notation}
\newtheorem{problem}{Problem}
\newtheorem{prop}{Proposition}
\newtheorem{rem}{{\it Remark}}
\newtheorem{solution}{Solution}
\newtheorem{summary}{Summary}
\numberwithin{equation}{section}
\newenvironment{pf}[1][Proof]{\noindent{\it {#1.}} }{\ \rule{0.5em}{0.5em}}
\newenvironment{ex}[1][Example]{\noindent{\it {#1.}}}

\thispagestyle{empty}


\begin{center}

{\LARGE\scshape Nonsingular Rotating Black Holes in the Dark-Energy Dominated Universe \par}
\vskip15mm

\textsc{Ram\'{o}n Torres\footnote{E-mail: ramon.torres-herrera@upc.edu}}
\par\bigskip
{\em
Dept. de F\'{i}sica, Universitat Polit\`{e}cnica de Catalunya, Barcelona, Spain.}\\[.1cm]
%


\end{center}

\begin{abstract}
Motivated by quantum-gravity scenarios that replace the classical black hole singularity with a regular core, and by the possibility that the dark-energy sector may be scale dependent, we construct a broad class of nonsingular rotating black-hole spacetimes embedded in an improved de Sitter--like background with either constant or running $\Lambda$. Because the Newman--Janis algorithm is generically incompatible with a cosmological-constant fluid, we instead propose a generalized Kerr--Schild construction on a (possibly scale-dependent $\Lambda$) de Sitter seed, yielding a Carter-type metric characterized by a mass function and a $\Lambda$ function. Our construction provides a direct map from static, spherically symmetric regular models to their rotating counterparts.
We derive sharp regularity conditions at the ring
and we identify a minimal-order subclass. We analyze chronology and show that, for non-negative mass function and $\Lambda$ above a certain negative limit, the spacetimes are stably causal.
For minimal-order geometries with non-negative mass, we prove that the weak energy condition must be violated.
Finally, we illustrate the framework with an asymptotic-safety--inspired model
and discuss horizon structure, surface gravities, and conformal diagrams. 
These results provide a controlled stationary-axisymmetric framework for future numerical studies of geodesics, shadows and horizon-scale observables of regular rotating black holes in dark-energy backgrounds.
\end{abstract}


\setcounter{equation}{0}

\section{Introduction}

According to our present understanding —grounded primarily in General Relativity (GR)— when an astrophysical object’s self-gravity overwhelms all available sources of pressure support, dynamical collapse forms 
a black hole. Because realistic progenitors generically possess nonzero total angular momentum, conservation of angular momentum implies a spinning remnant; after transient losses through neutrino emission, electromagnetic outflows, and gravitational-wave radiation, the exterior vacuum relaxes, by the black-hole uniqueness results under standard assumptions, to a stationary black hole geometry. Thus, within GR, astrophysical black holes are expected to be rotating black holes (RBHs).

Classical black-hole solutions of general relativity typically exhibit an internal curvature singularity, signaling a breakdown of the classical description. Many authors, therefore, regard the appearance of singularities less as a faithful prediction of nature and more as evidence that the theory is being extrapolated beyond its domain of validity. Motivated by this viewpoint, a substantial body of work has developed singularity-free (regular) black-hole models using complementary effective approaches (see, for example, \cite{A-BI,A-BII,B&V,Bardeen,A&B2005,B&R,Frolov2014,G&P2014,Hay2006,H&R2014,dust2014} and references therein).


With regard to the background of black holes in the Universe, the standard cosmological model, $\Lambda$CDM, assumes the existence of a strictly constant cosmological term $\Lambda$ in Einstein’s equations (which is responsible for the late--time accelerated expansion of the Universe).
Nevertheless, 
a broad body of work has also explored the possibility of evolving dark-energy using SNe Ia, BAO, CMB, and $H(z)$ measurements. Global analyses generally conclude that dynamical models are not excluded and sometimes mildly preferred, especially when including LSS and bispectrum information \cite{Avsajanishvili2022}.
Moreover, 
recent DESI DR1 BAO measurements provide a dense set of distance indicators over
$0.1\lesssim z\lesssim 3.5$ \cite{DESI2024VI}. While BAO-only constraints remain
broadly consistent with $\Lambda$CDM, joint analyses with Planck 2018 CMB and
type-Ia supernova compilations (e.g.\ Pantheon+) can favor dynamical dark energy (DDE)
in the $(w_0,w_a)$ framework, with best-fit regions typically drifting toward
$w_0>-1$ and $w_a<0$ and the cosmological-constant point $(w_0,w_a)=(-1,0)$
disfavored at the $\sim 2.5$--$4\sigma$ level in some dataset/priors choices
\cite{DESI2024VI,Berghaus2024DESI,Roy2024DESI}. Non-parametric reconstructions
using BAO+CMB+SNe similarly report mild departures from a strictly constant
dark-energy density over limited redshift ranges \cite{Calderon2024Crossing,Jiang2024},
although the robustness of these hints remains under scrutiny
\cite{Roy2024DESI,Chudaykin2024,Orchard2024}. Taken together, current data
do not rule out $\Lambda$CDM, but they 
motivate exploring theoretically controlled
scenarios in which the effective cosmological term is not strictly constant.


On the other hand, from a theoretical point of view, the cosmological constant raises severe theoretical problems: the enormous discrepancy between the observed $\Lambda$ and naive quantum-field-theoretic estimates (the \textit{cosmological constant problem}) and the \textit{coincidence problem} of why the vacuum and matter densities are of the same order precisely today \cite{Weinberg1989,Carroll2001}. 
These difficulties motivate scenarios in which the dark energy sector may have dependence on the spacetime point considered. 
Broadly speaking, dynamical dark energy models extend $\Lambda$CDM in several directions (for comprehensive reviews see, e.g., \cite{Copeland2006DEReview,Avsajanishvili2022}):
Quintessence and k-essence \cite{CaldwellLinder2005,Linder2008,Tsuji}, Phantom Scenarios and Model-Independent Parametrizations \cite{Bahamonde,Avsajanishvili2022}, Interacting dark energy \cite{Wang}, Modified gravity \cite{Nojiri,Chudaykin2024},
Running vacuum and running cosmological constant \cite{ShapiroSola2009,SolPeracaula2022RSTA,MorenoPulido2023EPJC}
In particular, let us highlight running-vacuum models which provide a theoretically motivated realization of dynamical dark
energy in which the vacuum term $\Lambda$ becomes an effective quantity that can vary with
cosmological curvature scales, as suggested by QFT in curved spacetime and semiclassical
gravity. Recent global fits to pre- and post-DESI-era combinations of SNIa, BAO, $H(z)$, LSS
and CMB data indicate that a small level of vacuum dynamics,
$|\nu_{\rm eff}|\sim 10^{-3}\!-\!10^{-2}$, can improve the overall likelihood relative to
$\Lambda$CDM and may help ease the $\sigma_8$ and $H_0$ tensions, though the robustness of
these indications are under active scrutiny
\cite{SolPeracaula2022RSTA,MorenoPulido2023EPJC,SolaPeracaula2023RVMStatus,DESI2024VI,Berghaus2024DESI}.


\subsection{The Newman-Janis algorithm and Dark Energy}

In 1965, Newman and Janis \cite{N&J} discovered that it was possible to obtain Kerr's solution by applying an algorithm to a spherically symmetric and static \emph{seed} metric: Schwarzschild's solution. A significant step towards understanding the algorithm, its possibilities, generalizations and limitations was carried out in \cite{Drake&Szek}. The generalized algorithm allows one to take any static spherically symmetric seed metric and obtain a rotating axially symmetric offspring from it. The application to regular RBH followed \cite{B&M}: One starts with a regular, static, and spherically symmetric black hole derived from a specific theoretical framework or adopted as a phenomenologically motivated effective model. Then, one applies the generalized Newman-Janis (N-J) algorithm to try to construct a regular RBH. 

Unfortunately, the \textit{offspring} has different geometrical properties and also different physical properties. For example, the seed metric can be a perfect fluid, but the offspring will never be another perfect fluid \cite{Drake&Szek}. In this way, if one takes a de Sitter (or anti-de Sitter) spacetime, which can be interpreted as containing a perfect fluid (with density and pressure related by $p=-\rho$), and one applies the algorithm, one does not obtain a solution for a rotating black hole in a background with a cosmological constant ($\mathbf T=-\frac{\Lambda}{8\pi} \mathbf g$), i.e., Carter's solution, but a \textit{rotating imperfect fluid} \cite{A-A}. In fact, as analyzed in detail in \cite{BelGon}, the resulting spacetime is not physically a $\Lambda$-term: its stress tensor
changes from the nonrotating de Sitter Segre type [(111,1)] to [(11)(1,1)] --and therefore does not reproduce the physical content of Carter’s Kerr–de Sitter geometry-- and the null energy condition is generically violated. Hence Newman–Janis-type constructions are not reliable for generating rotating black holes in a cosmological-constant background.

\subsection{Goal and structure of the article}

Even if a regular RBH model comes from an approach to Quantum Gravity Theory, in this article we will assume that it can be reasonably well described by a manifold endowed with its corresponding metric. 
Our aim is to find the general metric for a nonsingular rotating black hole embedded in a spacetime with dark energy modelled as a running (or constant) $\Lambda$ --- with $\Lambda$ acting as an effective local description tied to the curvature/scale of the spacetime. Our metric will accommodate the running $\Lambda$ consistently, without the artifacts introduced by the N-J algorithm. We will list the conditions required for a good chronological behaviour, analyze the fulfillment of energy conditions and describe its causal structure.  
We will also propose an alternative algorithm which will provide us with the spacetime geometry for stationary and axially symmetric black holes from static spherically symmetric black hole spacetimes.

This article is organized as follows.
In Sec.\ref{secRRBH} we introduce a generalized Kerr--Schild construction with a (possibly scale-dependent) dark-energy sector, and derive a stationary, axisymmetric metric describing a rotating black hole embedded in a background characterized by a running (or constant) cosmological term.
In Sec.\ref{SCS} we formulate and prove the requirements needed to avoid scalar curvature singularities, obtaining necessary conditions on the differentiability and core behavior of the mass function and the running cosmological term.
In Sec.\ref{SSTRBH} we prescribe a systematic procedure to promote a given static, spherically symmetric (regular) black-hole model
to its rotating counterpart within the class constructed in Sec.\ref{secRRBH}.
In Sec.\ref{BLF} we derive the Boyer--Lindquist-like representation of the metric 
thereby making the relevant hypersurfaces and symmetry structure manifest.
In Sec.\ref{chrono} we analyze the chronological properties of the resulting spacetimes, establishing sufficient conditions for stable causality in a natural block decomposition.
In Sec.\ref{SLSWEC} we study the stress--energy tensor 
and show that, for minimal-order regular spacetimes with non-negative mass function, the weak energy condition must be violated.
In Sec.\ref{KHSG} we characterize the Killing horizons, 
compute the associated horizon angular velocity and obtain an explicit expression for the surface gravity.
In Sec.\ref{SNSRBHM} we present an explicit nonsingular rotating black-hole model motivated by asymptotic safety via a renormalization-group-improvement prescription, and we discuss its horizon structure and global causal diagram in representative parameter regimes.
Finally, Sec.\ref{Conclu} summarizes the main results and open problems.

\section{Metric for a regular RBH embedded in Dark Energy}\label{secRRBH}

In order to obtain the metric suitable for a regular RBH embedded in dark energy, we will introduce a generalized Kerr-Schild ansatz \cite{B&G,Kramer} in which the spacetime metric $\tilde g$ for our spacetime $\tilde V$ is derived from a seed metric $g$ of a seed spacetime $V$. The metrics will be related through the relation
\begin{equation}\label{KSm}
\tilde g_{\alpha\beta}= g_{\alpha\beta}+ H l_\alpha l_\beta,
\end{equation}
where both metrics are non-flat, $H$ is a scalar function and $\mathbf l$ is a null vector with respect to both metrics. A list of some properties of the generalized Kerr-Schild transformation can be found in \cite{B&G}\cite{Kramer}.

Note that this procedure was used in the pioneering work by Carter \cite{Carter73} to obtain a rotating black hole from a seed de Sitter (or AdS) spacetime ($T_{\alpha\beta}=-\Lambda/(8\pi)\, g_{\alpha\beta}$).   
This procedure has also been applied recently in the presence of a cosmological constant in \cite{GLPP} and \cite{N&S}. Nevertheless, the specific procedure in these references is not fully suitable for possibly varying dark energy. 

In the general case considered here, our seed metric will be a spacetime containing dynamical dark energy. (The cosmological constant case will be treated as a particular case of the general one). This will be achieved by starting with a de Sitter-like spacetime and allowing $\Lambda$ to depend on the characteristic scale at which it is probed (and some parameters depending on the chosen approach to dark energy) \cite{W&K,Weinberg79,Reuter98}.
The line-element for the de Sitter-like spacetime in Carter's coordinates $\{\widetilde t,r,\theta, \widetilde \varphi\}$ takes the form \cite{Carter73}:

\begin{equation}\label{ImpDSitter}
    \begin{aligned}
d s_{0}^{2}=&  \Sigma \left\{\frac{d r^{2}}{\left(a^{2}+r^{2}\right) \Psi }+\frac{d \theta^{2}}{\Delta_\theta}\right\} 
 +\sin ^{2} \theta\frac{\Delta_\theta}{\Sigma \Xi^2 }
\left[a d \widetilde t-\left(a^{2}+r^{2}\right) d \widetilde\varphi\right]^{2} -\\
& -\frac{\Psi \left(a^{2}+r^{2}\right)}{\Sigma \Xi^2} \left(d \widetilde t-a \sin ^{2} \theta d \widetilde \varphi\right)^{2},
\end{aligned}
\end{equation}

where we have defined 
\begin{eqnarray*}
    \Sigma&\equiv& r^2+a^2 \cos^2\theta,\\
    \Psi&\equiv& 1- \frac{\Lambda r^2}{3},\\ 
    \Xi&\equiv& 1+\frac{\Lambda a^2}{3},\\ 
    \Delta_\theta&\equiv& 1+\frac{\Lambda a^2 \cos^2\theta}{3}.
\end{eqnarray*}

Note that, for constant $\Lambda$, (\ref{ImpDSitter}) is simply ordinary de Sitter spacetime written in Carter’s twisted coordinates (in which the parameter $a$ has only a coordinate meaning); by the untwisting transformation given by Carter (\cite{Carter73}, eqs. (5.68)–(5.75)) it can be brought to the familiar static-patch form.

On the other hand, in the case of applying the Kerr-Schild ansatz to obtain a rotating black hole it is convenient to write the scalar function $H$ as
\[
H=\frac{2 \mathcal M r}{\Sigma},
\]
where the scalar function $\mathcal M$ is defined as the \textit{mass function}.

We choose the null vector $\mathbf l$ in Carter form \cite{Carter73}:
\begin{equation}\label{lfC}
l_\alpha dx^\alpha \equiv \frac{d \widetilde t-a \sin^2 \theta d\widetilde\varphi}{\Xi}+
\frac{\Sigma dr}{\Psi (r^2+a^2)},
\end{equation}
which is null with respect to the seed metric (and, thus, to the generated metric) and will provide the required symmetries for $\tilde g$. In this way, from (\ref{KSm}),
the line element derived from the improved de Sitter spacetime will be

\begin{equation}\label{RBHgeneral}
 \begin{aligned}
ds^{2}=&  \Sigma \left\{\frac{d r^{2}}{\left(a^{2}+r^{2}\right) \Psi }+\frac{d \theta^{2}}{\Delta_\theta}\right\} 
 +\sin ^{2} \theta\frac{\Delta_\theta}{\Sigma \Xi^2 }
\left[a d \widetilde t-\left(a^{2}+r^{2}\right) d \widetilde\varphi\right]^{2} -\\
& -\frac{\Psi \left(a^{2}+r^{2}\right)}{\Sigma \Xi^2} \left(d \widetilde t-a \sin ^{2} \theta d \widetilde \varphi\right)^{2}+
\frac{2 \mathcal M r}{\Sigma} \left(\frac{d \widetilde t-a \sin^2 \theta d\widetilde\varphi}{\Xi}+
\frac{\Sigma dr}{\Psi (r^2+a^2)}\right)^2,
\end{aligned}
\end{equation}

where now $a$ acquires a physical interpretation as a rotation parameter that measures the angular momentum per unit of mass \cite{gravit}. Meanwhile, the scalar functions  $\mathcal M$ and $\Lambda$ will depend on both the characteristic scale $k$ at which they are probed (and some parameters depending on the chosen theoretical approach) \cite{W&K,Weinberg79,Reuter98}. In this curved spacetime the relevant physical coarse-graining scale is not global and varies from event to event, motivating a local identification $k=k(x)$. In this way, $\mathcal M$ and $\Lambda$ will be functions of the spacetime point considered. However, in order for the spacetime to be stationary and axially symmetric, as we want for our class of rotating black holes, we should impose that the scalar functions $\mathcal M$ and $\Lambda$ can only depend on the coordinates $r$ and $\theta$, so that $\partial_{\widetilde t}$ and $\partial_{\widetilde \varphi}$ will be killing vectors associated with the required symmetries\footnote{Note now that the choice of $\mathbf l$ makes it 
invariant under the killing vectors $\partial_{\widetilde t}$ and $\partial_{\widetilde\varphi}$, which is key for the derived spacetime to be stationary and axially symmetric.}.

In the literature, regular spinning black holes with nontrivial angular dependence have been explored in the locality-principle constructions of Eichhorn and Held \cite{E&H}, with subsequent analysis showing that these geometries are generically \textit{non-circular} \cite{DEH,Chen2022}; further phenomenological developments of this class were given in \cite{E&H2023}. A conceptually related asymptotic-safety-inspired construction of quantum-improved regular Kerr black holes was discussed in \cite{Chen}. In those models, allowing angular dependence can obstruct the existence of a well-defined entropy and may even lead to singular behavior at the horizon.

In this article we will focus in the case in which $\mathcal M$ and $\Lambda$ only depend on $r$.
This restriction 
is not intended as a claim of full generality, but as a controlled truncation: 
it preserves analytic tractability and, crucially, within the generalized Kerr–Schild family considered here it is the condition under which the Frobenius integrability required for the Boyer–Lindquist-like form of Sec.\ref{BLF} holds. (The results of \cite{DEH,Chen} provide additional motivation for this restriction, even though they do not by themselves constitute a general theorem for the full class of metrics studied here).


Let us remark that, unlike in the Kerr-de Sitter case, in the case with a varying $\Lambda$, $\Xi$ cannot be set to 1 by rescaling $\widetilde t$ and $\widetilde \varphi$.

The hypersurfaces $r=$constant have as their normal $\mathbf n=dr\ (\Leftrightarrow n_\mu=\delta^r{}_\mu)$ and therefore $\mathbf n ^2=n_\alpha n_\beta g^{\alpha\beta}=g^{rr}$. For the metric (\ref{RBHgeneral}) this can be written as
\[
\mathbf n ^2=g^{rr}=\frac{\Delta_r}{\Sigma},
\]
where we define $\Delta_r$, the numerator of $g^{rr}$, as
\[
\Delta_r\equiv -\frac{\Lambda(r)}{3} r^2 (r^2+a^2)+r^2-2 \mathcal M(r) r +a^2.
\]
It can be verified that there are no problems in treating the $r=$constant hypersurfaces in Carter's coordinates regardless of the value of $\Delta_r$.
In this way, $r=$constant is a spacelike hypersurface if $\Delta_r>0$, timelike if $\Delta_r<0$ and lightlike if $\Delta_r=0$ 
\footnote{In the general case treated here one could theoretically have a region with $r_{min}\leq r\leq r_{max}$ where all $r=$constant hypersurfaces could be lightlike if the functions $\mathcal M$ and $\Lambda$ 
satisfy in that region the relation
\[
\mathcal M =\frac{a^2+r^2}{2 r} \left( 1-\frac{\Lambda}{3} r^2\right).
\]
Clearly, in these regions either $\Lambda$ and/or $\mathcal M$ have to be chosen on purpose to get this behaviour. In this article, these regions will not be considered since there is no physical argument for their existence.}.

\section{Avoiding scalar curvature singularities}\label{SCS}

We say that there is a \emph{scalar curvature singularity} in a spacetime if any scalar invariant polynomial in the Riemann tensor diverges when approaching it along any incomplete curve. We want our seed de Sitter-like spacetime with metric (\ref{ImpDSitter}) to be devoid of singularities in order for it to be able to generate a singularity-free spacetime. Clearly $\Psi=0$ is a harmless coordinate singularity, but $\Sigma=0$ $\Leftrightarrow (r=0,\theta=\pi/2)$ could be a scalar curvature singularity. In order to prevent it we have the following
\begin{prop}\label{reg_I-DS}
  Assuming an improved de Sitter metric (\ref{ImpDSitter}) possesses a $C^4$ function $\Lambda(r)$, a necessary and sufficient condition for its curvature invariants to be finite at $\Sigma=0$ is
\begin{equation*}
\Lambda'(0)=\Lambda''(0)=\Lambda'''(0)=0.
\end{equation*}  
\end{prop}

\noindent\textbf{Proof.} This can be shown by evaluating the curvature (Ricci) scalar $\mathcal R$ near $(r=0,\theta=\pi/2)$ through a Taylor expansion of $\Lambda$. In the expansion one finds terms of the form $\Lambda'(0)/r^2,\ \Lambda''(0)/r^2$ and $\Lambda'''(0)/r$. Therefore, the curvature scalar will only be finite at the ring if $\Lambda'(0)=\Lambda''(0)=\Lambda'''(0)=0$. If we define $\xi\equiv a \cos\theta/r$ and $\tilde \xi$ as its limiting value along a path approaching $(r=0,\theta=\pi/2)$, the curvature scalar tends to the finite values

\[
\mathcal R\rightarrow
\frac{4 a^{2} \left(\tilde \xi^{2}+1\right) \Lambda(0)^{2}+12 \left(\tilde \xi^{2}+1\right) \Lambda(0)+2 a^{4} \Lambda^{(4)}(0)}{\left(\Lambda(0) \,a^{2}+3\right) \left(\tilde \xi^{2}+1\right)},
\]
for finite $\tilde\xi$, and 
\[
\mathcal R\rightarrow 4 \Lambda(0)
\]
for $\tilde\xi$ infinite.
Now, it suffices to examine the finiteness of a full set of invariants like the Carminati-McLenaghan invariants \cite{C&ML} around $(r=0,\theta=\pi/2)$ with the found conditions for non-singularity to verify that the conditions are in fact necessary and sufficient. The calculations were computed using Maple. \hfill $\Box$

Of course, our goal is to get the conditions that a spacetime for the rotating black hole with metric (\ref{RBHgeneral}) should satisfy to be devoid of scalar curvature singularities.

\begin{theorem}\label{condisreg}
  Assuming a RBH metric (\ref{RBHgeneral}) possessing a $C^3$ function $\mathcal M(r)$ and a $C^4$ function $\Lambda(r)$, a necessary condition for its curvature invariants to be finite at $(r=0,\theta=\pi/2)$ is
\begin{eqnarray*}
 \mathcal M (0)&=& \mathcal M' (0)= \mathcal M'' (0)=0 .\\
 \Lambda'(0)&=&\Lambda''(0)=\Lambda'''(0)=0.
\end{eqnarray*}  
\end{theorem}

A metric for a rotating BH satisfying the conditions in the Theorem will be called \textit{scalar-regular} (S-R) metric.

\noindent\textbf{Proof.} By evaluating the curvature (Ricci) scalar \footnote{Let us remark that the expressions of the all the invariants for this family of spacetimes are extremely complex and lengthy, so that we will only make explicit their expressions when the ring is approached.} $\mathcal R$ near the ring through a Taylor expansion of $\mathcal M$ and $\Lambda$ one finds terms of the kind $\Lambda'(0)/r^2,\ \Lambda''(0)/r^2,\ \Lambda'''(0)/r,  \mathcal M'(0)/r^2$ and $\mathcal M''(0)/r$. Therefore, the curvature scalar will only be finite at the ring if $\Lambda'(0)=\Lambda''(0)=\Lambda'''(0)=\mathcal M'(0)=\mathcal M''(0)=0$. If we define $\xi\equiv a \cos\theta/r$ and $\tilde \xi$ as its limiting value along a path approaching the ring, the curvature scalar at the ring tends to the finite values

\[
\mathcal R\rightarrow\frac{4 a^2 (\tilde \xi^2+1) \Lambda(0)^2+4 ( \mathcal M'''(0) a^2+3 \tilde\xi^2+3) \Lambda(0)+2 \Lambda^{(4)}(0) a^4 +12 \mathcal M'''(0) }{(\Lambda(0) a^2+3)(\tilde\xi^2+1)},
\]
for finite $\tilde\xi$, and 
\[
\mathcal R\rightarrow 4 \Lambda(0)
\]
for $\tilde\xi$ infinite. \footnote{Note that, in the absence of $\Lambda$, one gets the known results \cite{TorresReg}
$\mathcal R \rightarrow 4 \mathcal M'''(0)/(\tilde\xi^2+1)$ for finite $\tilde\xi$ and $\mathcal R \rightarrow 0$ for infinite $\tilde\xi$.}

On the other hand, if one evaluates the Kretschmann scalar $\mathcal K\equiv R_{\alpha\beta\gamma\delta} R^{\alpha\beta\gamma\delta}$ near the ring, one finds a term that behaves as $\mathcal M(0)/r^6$. Therefore, to avoid scalar curvature singularities we have to also impose $\mathcal M(0)=0$. If the conditions in the theorem are fulfilled, then the  Kretschmann scalar takes finite values when the ring is approached. Its limiting expression is too long to be shown here, but it can be verified in the appendix.  \hfill $\Box$

The soundness of the theorem is confirmed by the Carminati-McLenaghan invariant $r_1$ (quadratic in the trace-free Ricci tensor: $r_1 = \frac{1}{4} S_{ab} S^{ab}$, where $S_{ab} = R_{ab} - \frac{1}{4} R g_{ab}$) \cite{C&ML} that, under the assumptions of the theorem takes the finite expression

\begin{equation}
\begin{aligned}
r_{1} \;\rightarrow\;
\frac{1}{
      \bigl(\Lambda(0)\,a^{2}+3\bigr)^{2}\,
      (\tilde{\xi}^{2}+1)^{4}}
\Bigl[
  &\frac{a^{8}\bigl(\tilde{\xi}^{2}+\tfrac19\bigr)\,(\tilde{\xi}^{2}+1)\,{\Lambda^{(4)}}(0)^{2}}{8}
\\[3pt]
  &+\frac{(\tilde{\xi}^{2}-\tfrac13)\,a^{4}\,\mathcal M'''(0)\,\tilde{\xi}^{2}\,
          \bigl(\Lambda(0)\,a^{2}+3\bigr)\,\Lambda^{(4)}(0)}{2}
\\[3pt]
  &+\mathcal M'''(0)^{2}\,\tilde{\xi}^{4}\,
     \bigl(\Lambda(0)\,a^{2}+3\bigr)^{2}
\Bigr].
\end{aligned}
\end{equation}

for finite $\tilde{\xi}$ and $r_1\rightarrow 0$ for infinite $\tilde{\xi}$. \footnote{Again, in the absence of $\Lambda$, one gets, as a subcase, the known results \cite{TorresReg}
$r_1 \rightarrow \mathcal M'''(0)^2 \tilde{\xi}^4/(\tilde{\xi}^2+1)^4$ for finite $\tilde{\xi}$ and $r_1 \rightarrow 0$ for infinite $\tilde{\xi}$.}

In the same way, the complex Carminati-McLenaghan invariant $w_1$ (which measures the holomorphic self-dual contraction of the Weyl tensor:
$w_1 = \frac{1}{8} \left( C_{abcd} + i \, {}^*C_{abcd} \right) C^{abcd}$) \cite{C&ML} is finite if the conditions of the theorem are fulfilled.  Its limiting expression is too long to be shown here, but it can be verified in the appendix.

Therefore, if the necessary differentiability conditions are fulfilled, the mass function and $\Lambda$ of a scalar-regular black hole should be, respectively, around $r=0$ of the form $\mathcal M(r) = O(r^n)$ with $n\geq 3$ and $\Lambda(r)-\Lambda(0)=O(r^m)$ with $m\geq 4$. A usual case in the literature of regular black holes (without $\Lambda$) occurs when $\mathcal M(r) = M_3 r^3+ O(r^4)$ ($M_3\neq 0$). If, additionally, $\Lambda(r)-\Lambda(0)= \Lambda_4 r^4+O(r^5)$ ($\Lambda_4\neq 0$) we will call the spacetime a \emph{minimal-order spacetime}. 

It should be stressed that Theorem \ref{condisreg} guarantees the finiteness of the curvature invariants at the ring. However, in general, the limiting values as $(r=0,\theta=\pi/2)$ is approached, depend on the path through $\xi= a \cos\theta/r$, so a scalar-regular spacetime may still possess a nontrivial directional structure near the ring.

\section{From regular spherically symmetric BHs to regular rotating BHs}\label{SSTRBH}

Consider a static, spherically symmetric line element with areal radius $r$,
\begin{equation}\label{gsimesf}
ds^2
= -f(r)\,dt^2 + f(r)^{-1}\,dr^2
+ r^2\left(d\theta^2+\sin^2\theta\,d\phi^2\right),
\end{equation}
where we parametrize
\begin{equation}
f(r)=1-\frac{2M(r)}{r}-\frac{\Lambda(r)}{3}\,r^2,
\end{equation}
with $M(r)$ and $\Lambda(r)$ supplied by an underlying Gravitational Theory (General Relativity, Asymptotic Safety, string-inspired effective gravity, loop-inspired regular black-hole models and so on).

Then we obtain a consistent and computable alternative to Newman–Janis-type constructions for models with a $\Lambda(r)$ sector, within the generalized Kerr–Schild ansatz adopted here, via the following steps:
\begin{enumerate}
    \item Provided the spherically symmetric model has a function $\Lambda(r)$ satisfying the conditions in Theorem \ref{condisreg}, one identifies the function $\Lambda(r)$ for the spherically symmetric model with the function $\Lambda(r)$ for the line element (\ref{ImpDSitter}). In this way, one obtains the metric of a modified de Sitter-like spacetime  $V$ which will act as a seed for the RBH spacetime.
    \item Using $\Lambda(r)$, one chooses the Kerr–Schild 1-form $\mathbf l$ of Carter form (\ref{lfC}); it is null with respect to the seed metric $g$ and the spacetime metric $\tilde g$ by construction. 
    \item One identifies the mass of the spherically symmetric model as the \textit{mass function} ($\mathcal M=M(r)$), provided the spherically symmetric model has a function $M(r)$ satisfying the conditions in Theorem \ref{condisreg}. 
    \item One selects an angular momentum per unit mass $a$ for the rotating black hole and obtains a line element of the type (\ref{RBHgeneral}) by putting together the pieces of the previous steps. 
\end{enumerate}

In other words,

\begin{corollary}\label{coro1} 
For any choice of rotation parameter $a$, the stationary axisymmetric metric \eqref{RBHgeneral}
with the same functions $M(r)$ and $\Lambda(r)$ defines a rotating counterpart within our generalized
Kerr--Schild family.
In the limit $a\to 0$, the metric \eqref{RBHgeneral} reduces to the spherical line element
\eqref{gsimesf}.
Moreover, if $M$ and $\Lambda$ satisfy the near-ring conditions of Theorem \ref{condisreg}, then the resulting
rotating geometry is free of scalar polynomial curvature singularities at the ring.
\end{corollary}

In Sec. \ref{SNSRBHM}, we will apply this procedure to the asymptotic-safety–motivated nonsingular spherically symmetric solutions to construct corresponding families of regular rotating black-hole metrics.

\textit{Remark}: The construction developed here is adapted to static seeds in \textit{Schwarzschild gauge}, $g_{tt}g_{rr}=-1$; extending it to more general seeds with two independent radial metric functions is conceivable, and related Newman--Janis / Azreg-A\"inou-type approaches do exist in the literature \cite{BrahmaChenYeom2021,KocherlakotaNarayan2024,KocherlakotaNarayan2025}, but such generalizations require additional structure, need not preserve the seed metric exactly in the $a\to0$ limit, and can impose nontrivial constraints on the admissible matter sector \cite{KocherlakotaNarayan2024,KocherlakotaNarayan2025}.
We leave this broader generalization for future work.

\section{Boyer-Lindquist-like form}\label{BLF}

When working with general rotating black holes, it is customary to work with the metric expressed in Boyer-Lindquist-like (B-L) form. This is due to various advantages provided by this form: The symmetries are explicit, it has a near diagonal form (block-diagonalizes the $r$ sector), the key hypersurfaces are simple to identify and so on.

A remark is important here: In general, the B-L form cannot be obtained for $\mathcal M$ and $\Lambda$ depending both on $r$ and $\theta$. Frobenius Theorem implies that only in case we are working with a metric (\ref{RBHgeneral}) in which $\mathcal M=\mathcal M (r)$ and $\Lambda=\Lambda(r)$, it is possible to transform the line-element (\ref{RBHgeneral}) written in Carter coordinates to a Boyer-Lindquist-like form through a coordinate change $\{\widetilde t,r,\theta, \widetilde \varphi\} \rightarrow \{t,r,\theta, \varphi\}$\footnote{Note that the usual coordinate change \cite{GLPP}\cite{N&S} starting from \textit{spheroidal} coordinates to B-L-like coordinates cannot be used if $\Lambda\neq$ constant.} defined through

\begin{eqnarray*}
    d\widetilde t &=&dt+\frac{2 \mathcal M(r) r \Xi}{\Psi \Delta_r}\, dr,\\
    d\widetilde \varphi &=& d\varphi+ \frac{2 a \mathcal M(r) r \Xi}{(r^2+a^2) \Psi \Delta_r }\, dr.
\end{eqnarray*}

In this way, the line element of the RBH will be rewritten in the Boyer-Lindquist-like form as:
\begin{equation}\label{metricBL}
ds^2= \Sigma \left\{\frac{dr^2}{\Delta_r}+\frac{d\theta^2}{\Delta_\theta} \right\}+
\sin^2\theta \frac{\Delta_\theta}{\Sigma \Xi^2} (a dt-(r^2+a^2) d\varphi)^2-
\frac{\Delta_r}{\Sigma \Xi^2} (dt-a \sin^2\theta d\varphi)^2.
\end{equation}

As happens in Carter's coordinates (\ref{ImpDSitter}), in order for the angular sector to stay spacelike, $g^{\theta\theta}>0$ and this leads to the condition $\Delta_\theta>0$. In this way, a spacelike angular sector implies 
\begin{equation}\label{minLambda}
\Lambda>\Lambda_{min}\equiv \frac{-3}{a^2},
\end{equation}
which, for a varying $\Lambda$, can be interpreted as a limit on how negative $\Lambda$ can be in a model with fixed $a$.\footnote{Of course, for a given $\Lambda(r)$, one would prefer to interpret it as a limit in the rotation parameter.}

Some examples of well-known BHs that can be obtained by using this procedure are\footnote{Note that the last two solutions cannot be obtained using the generalized Newman--Janis algorithm.}: 
\begin{itemize}
    \item Schwarzschild metric ($\mathcal M=m=$ constant, $a=0$, $\Lambda=0$),
    \item Reissner–Nordström metric ($\mathcal M=m-q^2/(2r)$, $m=$constant, $q=$constant, $a=0$, $\Lambda=0$),
    \item Kerr metric ($\mathcal M=m=$ constant, $0<a^2\leq m^2$, $\Lambda=0$), 
    \item Kerr-Newman solution  ($\mathcal M=m-q^2/(2r)$, $m=$constant, $q=$constant, $a\neq 0$, $a^2+q^{2}\leq m^{2}$, $\Lambda=0$),
    \item Kerr-de Sitter (or Carter) solution  ($\mathcal M=m=$ constant, $a\neq 0$, $\Lambda>0$) and Kerr-anti-de Sitter solution  ($\mathcal M=m=$ constant, $a\neq 0$, $\Lambda<0$),
    \item Kerr-Newman-de Sitter solution  ($\mathcal M=m-q^2/(2r)$, $m=$constant, $q=$constant, $a\neq 0$, $\Lambda>0$) and Kerr-Newman-anti-de Sitter solution  ($\mathcal M=m-q^2/(2r)$, $m=$constant, $q=$constant, $a\neq 0$, $\Lambda<0$),
\end{itemize}

\section{Chronology}\label{chrono}

In a time-orientable spacetime, it is natural to require the absence of closed causal curves since the presence of such curves would generate familiar logical paradoxes. A spacetime with no closed causal curves is called \emph{causal} \cite{H&E}. If, moreover, this property survives under arbitrarily small perturbations of the metric, the spacetime is \emph{stably causal}.

It is well known that the standard maximal analytic extension of the Kerr solution is non-causal. Since the Kerr metric is a special case of (\ref{metricBL}), it is therefore natural to ask whether the maximal extensions of regular rotating black holes (RBHs) share this pathology.

As is evident from the transformation between Carter and Boyer–Lindquist (BL) coordinates, in the BL chart the hypersurfaces $r=r_H$ satisfying $\Delta_r\vert_{r_H}=0$ correspond to coordinate (not curvature) singularities, as documented in the standard subcases. Consequently, in BL coordinates one works with \emph{blocks}: we call a block \emph{type I} when $\Delta_r>0$ throughout the block, and \emph{type II} when $\Delta_r<0$.

\begin{prop}\label{proposchrono}
If $\Xi>0$ and $r \mathcal M(r)>0$ in a block of type I, then the block is stably causal. 
\end{prop}

\noindent\textbf{Proof.} First, note that $\Xi>0$ due to the constraints on $\Lambda$ (\ref{minLambda}). 
Let us consider the $t=$constant hypersurfaces. Their normal 1-form is $\mathbf{n}=dt$ and 
\begin{equation}\label{n2}
\mathbf n ^2=g^{tt}= - \frac{\Xi^2 F}{\Delta_\theta \Delta_r \Sigma}, 
\end{equation}
where 
\[
F\equiv (a^2+r^2)^2 \Delta_\theta - \sin^2\theta a^2 \Delta_r.
\]
Note, in particular that, if the specific block contains the ring ($r=0,\theta=\pi/2$) and there are no scalar curvature singularities then, as the ring is approached, $\mathbf n ^2 \rightarrow -(1+\Lambda(0) a^2 /3)^3<0$ ).

On the other hand, outside the ring, taking into account the non-negative functions appearing in expression (\ref{n2}), $t=$constant will be spacelike ($\mathbf{n}^2<0$)  if, and only if, $F>0$,
what can be written as 
\[
(a^2+r^2) \Sigma \Xi+\sin^2\theta a^2 r  \mathcal M>0.
\]

In this way, if $\Xi>0$ and the product $r \mathcal M(r)$ is non-negative the type I block will have a \textit{function} $t$ with $t=$constant spacelike hypersurfaces which, according to proposition 6.4.9 in \cite{H&E}, implies that the block will be stably causal.   $\Box$

In the absence of $\Lambda$, the condition of a non-negative $r \mathcal M(r)$ was already obtained in 
\cite{Maeda,Torres2023,torresSS}. Now we find that the limit (\ref{minLambda}) 
favors the absence of closed timelike curves. The need for this condition is expected since, as deduced from the comments in section \ref{BLF}, the violation of the condition would make the trajectory on a circle having $\theta(\tau)=\tau$, and $\phi(\tau)=\phi_0$ for
$\tau\in(0,\pi)$, while $\phi(\tau)=\phi_0+\pi$ (mod $2\pi$) for $\tau\in(\pi,2\pi)$ a timelike curve, which will be a \emph{closed} timelike curve. 

In the previous proposition (and through the use of $r \mathcal M(r)$) we are not excluding the possibility that the spacetime could be extended through the disk with $r<0$, which would provide a non-topologically trivial spacetime. From a historical point of view, this extension was needed for the singular Kerr solution to solve the problem of differentiability at the disk. Nevertheless, the non-topologically trivial spacetime introduced new problems: closed causal curves (clearly due to the existence of the $r<0$ regions) and the appearance of conical singularities. In the case treated in this article, where there are no scalar curvature singularities, there are no differentiability problems at the ring. Thus, the extension through the disk with $r<0$ can be avoided providing a \emph{topologically trivial spacetime} that, moreover, is free of all unphysical behaviours \cite{Torres2023}\cite{torresSS}. We can then write the following 

\begin{prop}\label{proposchrono2}
If $\Xi>0$ and $\mathcal M(r)>0$ in a block of type I of a nonsingular topologically trivial spacetime, then the block is stably causal. 
\end{prop}

\begin{prop}
Blocks of type II are stably causal. 
\end{prop}

\noindent\textbf{Proof.} For these blocks consider the $r=$constant hypersurfaces with normal 1-form $\mathbf n=dr$ satisfying
\[
\mathbf n ^2 = g^{rr}=\frac{\Delta_r}{\Sigma}\ <0.
\]
Therefore, there is a \textit{function} $r$ such that the $r=$constant hypersurfaces are spacelike. Then, according to proposition 6.4.9 in \cite{H&E}, type II blocks will always be stably causal.\hfill $\Box$

\section{Minimal-order spacetimes and the Weak Energy Condition}\label{SLSWEC}

The Zero Angular Momentum Observer (ZAMO) is defined by the condition that its conserved axial angular momentum per unit mass vanishes: $L_z=p_\varphi=g_{t\varphi} \dot t+ g_{\varphi\varphi} \dot \phi=0$, which implies the coordinate angular velocity,
\[
\Omega=\frac{d\varphi}{dt}=-\frac{g_{t\varphi}}{g_{\varphi\varphi}}=\frac{a [\Delta_\theta (r^2+a^2)-\Delta_r]}{F},
\]
Associated to these observers we can define the orthonormal tetrad:
\begin{eqnarray*}
    \hat e_t&=&\sqrt{\frac{\Xi^2 F}{\Delta_r \Delta_\theta \Sigma}} \left( \frac{\partial}{\partial t} +\Omega \frac{\partial}{\partial \varphi}\right),\\
    \hat e_r&=& \sqrt{\frac{\Delta_r}{\Sigma}} \frac{\partial}{\partial r},\\
    \hat e_\theta&=&\sqrt{\frac{\Delta_\theta}{\Sigma}} \frac{\partial}{\partial \theta},\\
    \hat e_\phi&=&\sqrt{\frac{\Xi^2 \Sigma}{F \sin^2\theta}} \frac{\partial}{\partial \varphi}.
\end{eqnarray*}

The effective density, radial pressure and angular pressures as measured by the ZAMO are defined through $\rho=T_{\mu\nu} e_t^\mu e_t^\nu$, $p_r=T_{\mu\nu} e_r^\mu e_r^\nu$, $p_\theta=T_{\mu\nu} e_\theta^\mu e_\theta^\nu$, $p_\varphi=T_{\mu\nu} e_\varphi^\mu e_\varphi^\nu$, respectively, where the effective energy-momentum tensor is $T_{\mu\nu}\equiv G_{\mu\nu}/(8\pi)$.

\begin{theorem}\label{TWEC}
    A rotating minimal-order spacetime with a non-negative mass function must violate the Weak Energy Condition.
\end{theorem}

\noindent\textbf{Proof.} To show this, we must take into account that the existence of a non-negative mass function for a minimal-order case ($M(r)= M_3 r^3+O(r^4)$ around $r=0$) requires $M_3>0$.
On the other hand, consider the ZAMOs located in the rotating axis ($\theta=0$). It can be verified that for these specific observers $T_{\mu\nu} e_a^\mu e_b^\nu=0$ if $a\neq b$, i.e., the effective energy-momentum tensor for them is diagonal (or \textit{Type I} \cite{H&E}). The WEC for type I spacetimes requires (among other inequalities) $\mbox{WEC}_1=\rho+p_r\geq 0$ and $\mbox{WEC}_2=\rho+p_\theta\geq 0$. Combining these inequalities one gets around $r=0$:
\[
\frac{\mbox{WEC}_1}{2}+\mbox{WEC}_2=-\frac{12 M_3}{a^2} r^2\geq 0,
\]
that demands $M_3<0$, in contradiction with a positive mass function around $r=0$. In this way, the WEC cannot be satisfied.\hfill $\Box$

Note that the violation of energy conditions is somehow expected since 
regularity requires evading at least one hypothesis of the singularity theorems; in effective models this commonly manifests as violation of (some) pointwise energy condition in the high-curvature region.

\section{Killing Horizons and surface gravity}\label{KHSG}

As we have seen in section \ref{secRRBH}, the isolated $r=r_H$ where $\Delta_r|_H=0$ are light-like hypersurfaces. These hypersurfaces are generated by the null geodesic congruence tangent to the Killing field \cite{Wald} 
\begin{equation}\label{defchi}
\vec\chi =\partial_t+\Omega_H \partial_\phi,
\end{equation}
with a specific constant angular velocity
\[
\Omega_H =\frac{a}{r_H^2+a^2}
\]
and $\vec\chi^2|_H=0$.

Therefore, using the standard definition, these null hypersurfaces are \textit{killing horizons}. The global behaviour of these horizons allows their classification as Cauchy horizons, event horizons or cosmological horizons. This requires the knowledge of the specific functions $\mathcal M(r)$ and $\Lambda(r)$. A particular case will be treated in section \ref{SNSRBHM}.

The surface gravity $\kappa$ in the horizon can be found through \cite{Wald} 
\[
\kappa^2=-\frac{1}{2} \nabla_\mu \chi_\nu\, \nabla^\mu \chi^\nu,
\]
which for our rotating black holes and using the normalization in (\ref{defchi}) is just
\[
\kappa=\frac{\Delta_r'(r_H)}{2 (r_H^2+a^2) \Xi_H},
\]
where the prime in $\Delta'$ stands for derivative with respect to $r$. Note that the surface gravity is constant in every horizon.

\section{Nonsingular RBH model from asymptotic safety}\label{SNSRBHM}

Weinberg’s asymptotic-safety scenario \cite{Weinberg} posits that quantum gravity may be ultraviolet complete due to a non-Gaussian renormalization-group (RG) fixed point, implying a predictive, scale-invariant regime near the Planck scale. In practical functional RG analyses, this idea is often explored within the Einstein–Hilbert truncation, where only the scale dependence of Newton’s coupling ($G(k)$) and the cosmological term ($\Lambda(k)$) is retained \cite{Reuter98}\cite{R&S}. To connect this truncated flow to black-hole spacetimes, one commonly adopts an “RG-improvement” prescription: the classical constants in a metric are replaced by running couplings evaluated at a position-dependent cutoff scale ($k=k(x)$) \cite{B&R}. While this procedure has some theoretical support \cite{B&P}, it is not unique and can generate unphysical singularities that reflect the improvement scheme rather than the underlying RG dynamics \cite{K&S}. For this reason, previous work \cite{AEP} has derived conditions under which RG improvement yields regular spherically symmetric geometries, focusing on the high-curvature regime where the flow is expected to approach scale invariance and the dynamics are governed by universal data such as critical exponents.

The Einstein-Hilbert subspace is spanned by the dimensionless running Newton coupling 
\[
g(k)\equiv G(k) k^2,
\]
and the dimensionless running cosmological \textit{constant}
\[
\lambda(k)\equiv \Lambda(k) k^{-2}.
\]

The expansion of their corresponding beta functions around a fixed point in the trans-Planckian regime $\{g_*,\lambda_*\}$ of the RG flow yields \cite{AEP}, to linear order and assuming real critical exponents\footnote{Of course, complex critical exponents are also possible \cite{AEP} and both cases have been found in the literature, but here we choose the real case to provide an example.}
\begin{eqnarray}
g(k)&=&g_*+g_1 \left(\frac{k}{M_P} \right)^{-\theta_1}+ g_2 \left(\frac{k}{M_P} \right)^{-\theta_2},\label{gk}\\
\lambda(k) &=& \lambda_*+\lambda_1 \left(\frac{k}{M_P} \right)^{-\theta_1}+ \lambda_2 \left(\frac{k}{M_P} \right)^{-\theta_2},\label{lk}
\end{eqnarray}
where $\theta_i$ are the critical exponents, $g_i$ and $\lambda_i$ are parameters and $M_P$ is Planck's mass.

In a spherically symmetric context, the RG-scale takes the form \cite{AEP}
\[
k(r)\simeq \tilde{\zeta} M_P r^{-\gamma},
\]
where $\tilde{\zeta}=\zeta (m G_0)^{\gamma-1} /M_P$ and $\zeta$ is a positive parameter.
By using the Renormalization Group improvement procedure it was found that the singularity resolution is possible if
\[
\gamma\geq 3/2,\ \ \lambda_*=0\ \ \mbox{and}\ \ \theta_i\geq 2
\]
are all satisfied.
Therefore, $\lambda(k)$ vanishes in the UV and it is dynamically generated by quantum fluctuations as the renormalization group flow departs from the scale-invariant Planckian regime. Consequently, the macroscopic value of $\Lambda$ observed at low energies emerges from the renormalization group flow as the energy scale $k$ is lowered, allowing for a physically realistic IR limit $\Lambda_0$.

On the other hand, the critical exponents derived in the literature (see, for example, \cite{BenedettiMachadoSaueressig2009,GiesKnorrLippoldt2015,DonaEichhornPercacci2014}) indeed satisfy $\theta_i \sim 2 \pm 0.5$.
In this way, the antiscreening character of quantum gravity fluctuations could lead to a sufficient weakening of the gravitational interaction at high curvature scales and might result in regular black holes.

In order to fix a particular model, in this article we choose a minimal model with parameters $\gamma= 3/2$ (which is the most straightforward choice based on physical arguments and dimensional analysis) and $\theta_i= 2$.
As a result, by using (\ref{gk}) and (\ref{lk}) , the values of the Newtonian coupling and the running $\Lambda$ in the UV limit will be, respectively
\begin{equation}\label{UVlimit}
 G(r)\simeq \frac{M_P}{m} r^3 \left[g_*+ \Gamma r^3\right] \text{and} \ \ \Lambda(r)\simeq \Lambda_{UV}=\text{constant,} 
\end{equation}
where $\Gamma$ is a constant.
On the other hand, in the IR limit, we should have 
\[ 
G(r) \rightarrow G_0\ \text{ and }\ \Lambda(r)\rightarrow \Lambda_0,
\]
the standard measured values. To get a model with the correct qualitative features we need interpolating functions between the UV limit and the IR limit. Probably one of the simplest examples could be
\begin{eqnarray}\label{model}
G(r)&=&\frac{G_0 r^3}{r^3+l^3},\nonumber\\
\Lambda (r)&=&\frac{l^2+\Lambda_0 r^4}{l^4+r^4},
\end{eqnarray}
where $l>0$ defines the scale separating the UV and the IR behaviour. 

$G(r)$ is qualitatively compatible with asymptotic safety (\ref{UVlimit}) since around $r=0$
\[
G(r)=\frac{r^3}{l^3} \left(1-\frac{r^3}{l^3} \right) G_0+ O\left(\frac{r^9}{l^9}\right)
\]
and that it satisfies the requirement $G(r\rightarrow\infty)=G_0$. The corresponding mass function is $\mathcal M(r)= M G(r)$, where $M$ is the mass of the BH at infinity, which satisfies the conditions for a scalar-regular RBH (theorem \ref{condisreg}). Note that this is precisely Hayward's mass function \cite{Hay2006}. 

On the other hand, $\Lambda (r)$ also satisfies the requirement $\Lambda(r\rightarrow\infty)=\Lambda_0$ and it is compatible with asymptotic safety (\ref{UVlimit}) and with the non-singularity condition for rotating black holes (theorem \ref{condisreg}) since around $r=0$
\[
\Lambda (r)=\Lambda_{UV} + (\Lambda_0-\Lambda_{UV}) \frac{r^4}{l^4}+ O(\frac{r^8}{l^8}),
\]
where $\Lambda_{UV}\equiv 1/l^2$. Note that, in this model if $l$ were of the order of Planck's length then $\Lambda_{UV}\sim 10^{121} \Lambda_0$. In this way, $\Lambda$ would have its naively \textit{expected} value \cite{Weinberg1989,Carroll2001} just in the interior of the black hole. However, it will decrease due to quantum fluctuations along the RG flow away from the scale invariant regime and towards its observed value in the IR regime.

By Collorary \ref{coro1},
a rotating black hole constructed using such functions:
\begin{itemize}
\item Satisfies the necessary conditions in Theorem \ref{condisreg} required to avoid scalar curvature singularities. Figures \ref{RP} and \ref{Kre} exemplify this by showing, respectively, the curvature scalar $\mathcal R$ and the Kretschmann scalar $\mathcal K$ around the ring.
\begin{figure}[ht]
\includegraphics[scale=0.9]{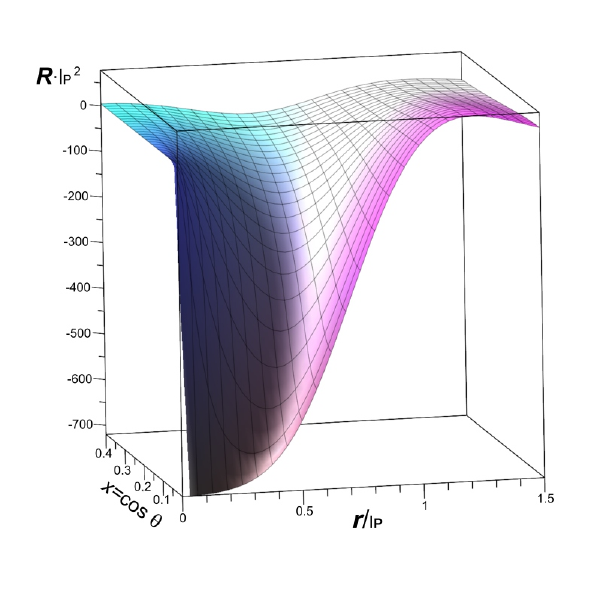}
\caption{\label{RP} Curvature scalar in the interior region around the ring, where one can explicitly see its finiteness and its directional behaviour at the ring, which corresponds with the one described in the text. For this particular plot the parameters have been chosen as: $l/l_P =1$, $M/M_P = 10$, $a/M_P = 5$, $\Lambda_0\cdot l^2 = 0.1$ which, at the ring, provide a maximum value for the curvature scalar $\mathcal R_{max}=4$ and a minimum value $\mathcal R_{min} \simeq -720.2857$.}
\end{figure}
\begin{figure}[ht]
\vspace{-0.5cm}
\includegraphics[scale=0.2]{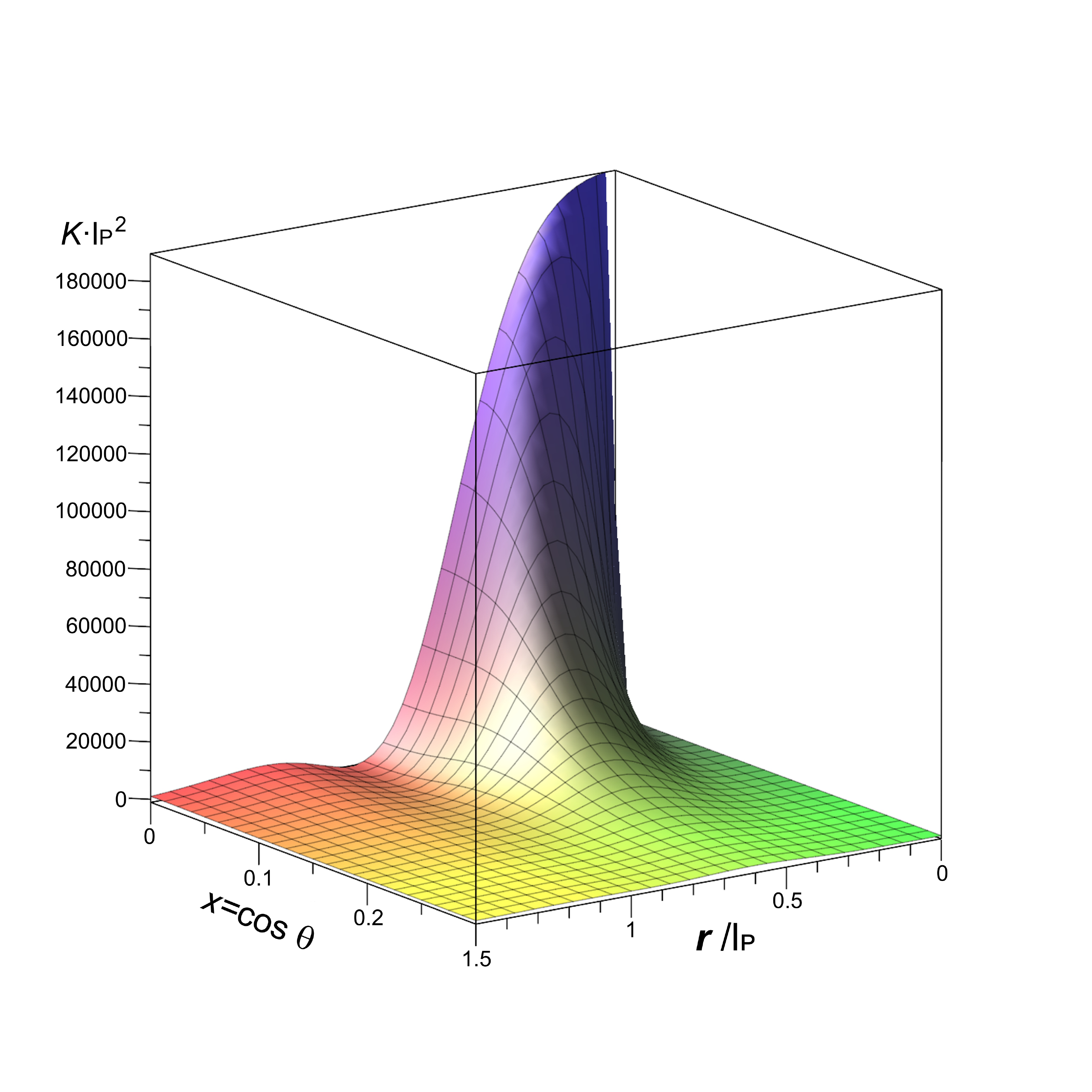}
\caption{\label{Kre} Kretschmann scalar in the interior region around the ring, where one can explicitly see its finiteness and its directional behaviour at the ring, which corresponds with the one described in the text and the expression given in the appendix. For this particular plot the parameters have been chosen as: $l/l_P =1$, $M/M_P = 10$, $a/M_P = 5$, $\Lambda_0\cdot l^2 = 0.1$ which, at the ring, provide a maximum value for the Kretschmann scalar $\mathcal K_{max} \simeq 189785$ and a minimum value $\mathcal K_{min} =8/3$.}
\end{figure}

\item Satisfies the conditions in proposition \ref{proposchrono2} for an asymptotically de Sitter spacetime ($\Lambda_0>0$), and therefore, all its blocks will be stably causal.

\item Is a \textit{minimal-order} spacetime with a non-negative mass function and has to violate the WECs according to Theorem \ref{TWEC}. Figure \ref{densplot} shows the effective density measured by ZAMO observers which is negative in a region close to the ring. Therefore, it is explicitly shown that the weak energy conditions are indeed violated.

\begin{figure}[ht]
\includegraphics[scale=0.25]{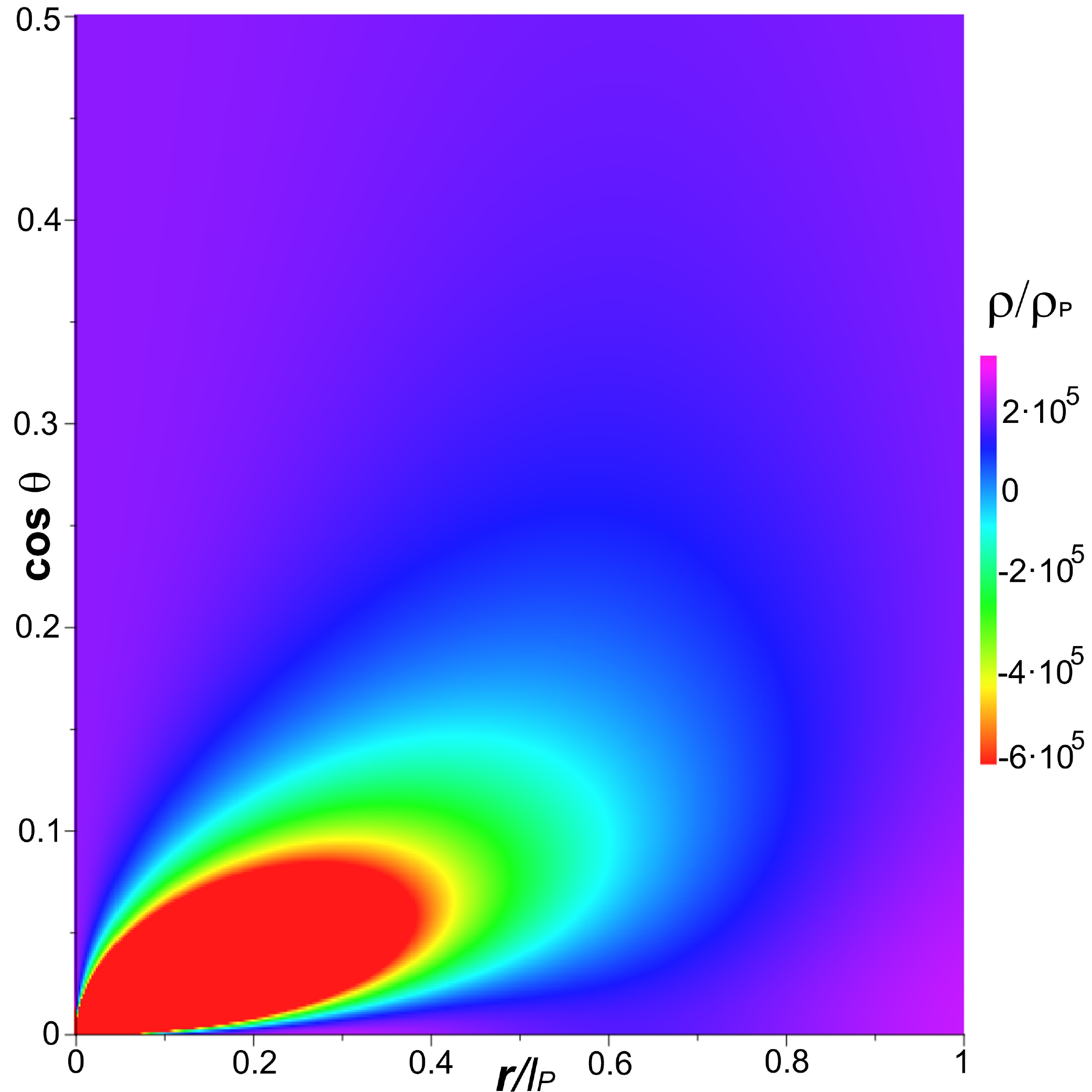}
\caption{\label{densplot} A heat map of the effective density as measured by ZAMO observers, where one can explicitly see that there is a region close to the ring where the effective density is mostly negative (red and surrounding non-blue regions). Therefore, the weak energy conditions are violated. For this particular plot the parameters have been chosen as: $l/l_P =1$, $M/M_P = 10$, $a/M_P = 5$, $\Lambda_0\cdot l^2 = 0.1$. The minimum value for the effective density is at $(r \simeq 4.0311 \cdot 10^{-4}, x =\cos \theta \simeq 5.9911 \cdot 10^{-4})$ where it takes the finite value $\rho\simeq -614682$.}
\end{figure}

\end{itemize}

\subsection{Horizons and causal structure}

As we have seen, the killing horizons for this family of spacetimes correspond with the isolated zeros $r=r_H$ of $\Delta_r$. It is not possible to find the exact analytical solutions for the zeros of $\Delta_r$. However, it is possible to find them numerically. 

Let us first consider the situation in which the effects of Quantum Gravity take place when $l\sim l_P$, the black hole is of astrophysical size ($M\gg M_P$), the angular momentum is small compared to the mass and the IR value of $\Lambda$ ($\Lambda_0$) takes the relatively small values measured in our Universe. 

\begin{figure}[ht]
\vspace{-1cm}
\includegraphics[scale=0.35]{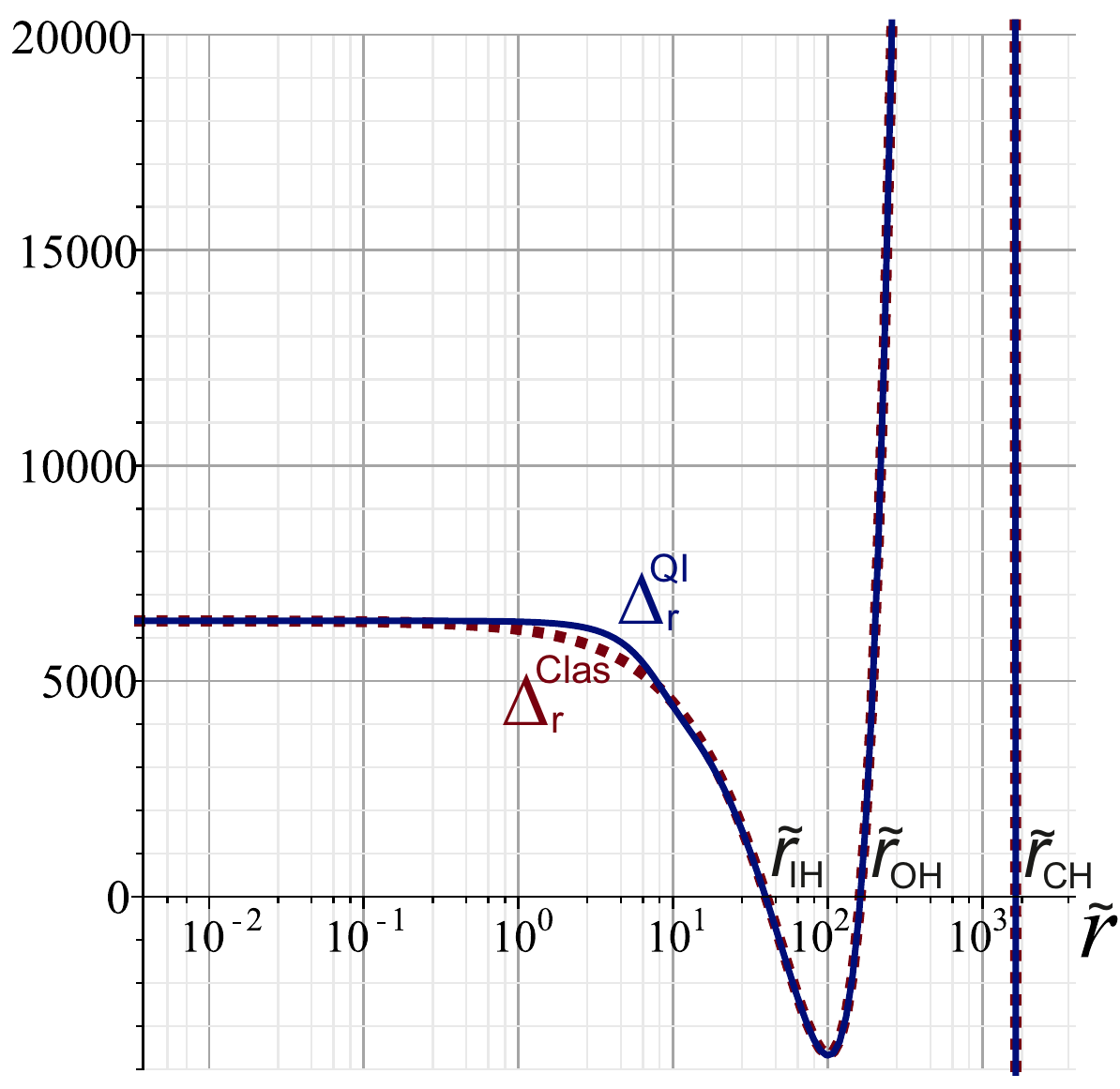}
\caption{\label{zeros} A plot of the zeros of $\Delta_r$ showing the position of the horizon for the classical (dashed brown) case versus the quantum improved case (solid blue). Note that to improve the visualization, a logarithmic scale has been used in the $r$ axes. Here $\tilde r\equiv r/l_P$. It can be verified that, even for $l=G_0 M/10$, the differences between the $\Delta_r$ functions are only appreciable in the quantum region below the inner horizon. Specifically, the displacement of the horizons was: $r_{IH}\simeq 0.991 r_{IH}^{Clas}$, $r_{OH}\simeq 1.002 r_{OH}^{Clas}$ and $r_{CH}\simeq 0.9999 r_{CH}^{Clas}$. The particular values used in this plot are $\{ l/l_P = 10, M/M_P = 100, a/M_P = 80, \Lambda_0 l_P^2 = 10^{-6}\}$.}
\end{figure}

For a classical Kerr-de Sitter black hole of this type, there are three positive horizons: Cauchy horizon (at $r=r_{IH}$), event horizon (at $r=r_{OH}$) and cosmological horizon (at $r=r_C$ ). In this situation, we expect the quantum corrections to the positions of these horizons to be negligible for the cosmological horizon (since $r_C\ggg l$),
very small for the event horizon (since $r_{OH}\gg l$) and 
even relatively small for the Cauchy horizon (since $r_{IH}>l$). Note however that, in the absence of a singularity, now there will not be a \textit{Cauchy horizon}, but an \emph{inner horizon}.
The positions of the horizons for a particular case are shown in figure \ref{zeros}.

\begin{figure}[ht]
\includegraphics[scale=0.8]{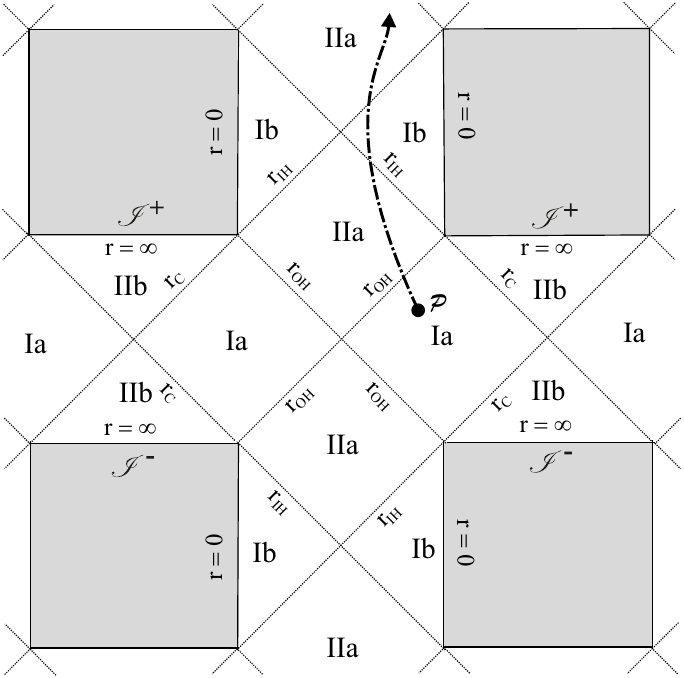}
\caption{\label{PenroseRRBH} A portion of the conformal diagram for the regular rotating black hole. The grey regions are not part of the spacetime. A test particle initially outside the black hole at point $\mathcal P$ can traverse the event horizon and will not reach a singularity. Instead, it could traverse the black hole, as shown. It could even reach the disk at $(r=0,\theta\neq\pi/2)$ and traverse it (keeping $r(\tau)\geq 0$, $\theta\rightarrow\pi-\theta$ and $\phi\rightarrow\phi+\pi$ (mod $2\pi$), i.e., staying in the Ib region) or reach the ring at $(r=0,\theta=\pi/2)$ , where there are no  singularities and continue its travel without any disruption.}
\end{figure}

The causal structure of a spacetime corresponding to such a BH is similar to the one corresponding to the classical Kerr-de Sitter solution. The main differences are in the interior of the black hole, where the singularity has been removed and the extension towards negative values of $r$ is no longer needed. The conformal diagram corresponding to such rotating regular BH is shown in figure \ref{PenroseRRBH}. The existence of horizons divides the figure in blocks of type I and II. We have also differentiated 
\begin{itemize}
\item Blocks of type Ia as those type I blocks between event horizons $r_{OH}$ and cosmological horizons $r_C$.
\item Blocks of type Ib as those type I blocks between $r=0$ and the inner horizon $r_{IH}$. 
\item Blocks of type IIa as those type II blocks between the event horizon $r_{OH}$ and the inner horizon $r_{IH}$ of the black hole.
\item Blocks of type IIb as those type II blocks between the cosmological horizon $r_C$ and infinity.
\end{itemize}

\begin{figure}[ht]
\vspace{-1cm}
\includegraphics[scale=0.4]{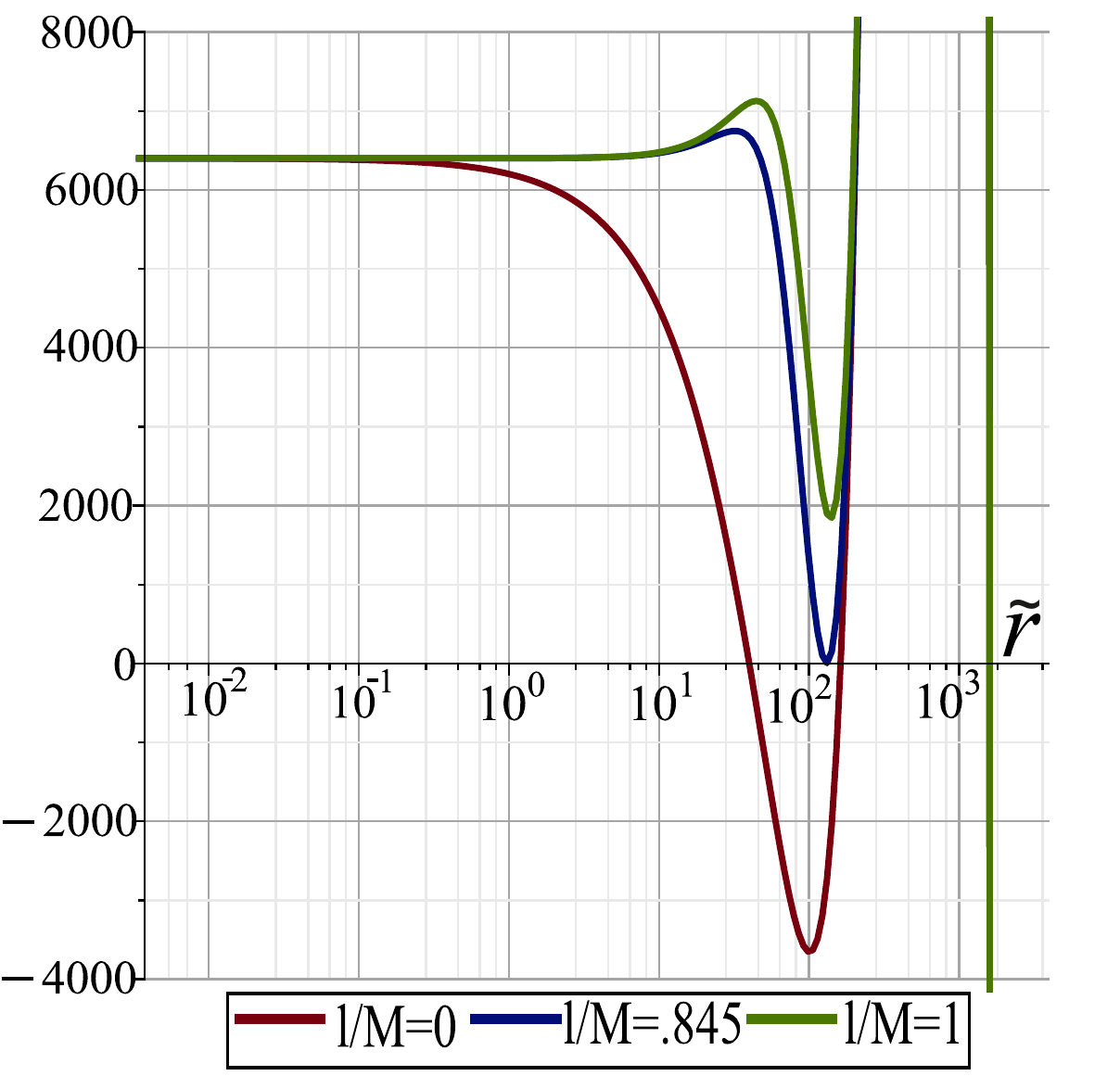}
\caption{\label{BigQuantumCH} A plot of the zeros of $\Delta_r$ for significant cases where big quantum corrections are allowed (blue and green curves) versus the classical case (brown curve). Here $\tilde r\equiv r/l_P$. In particular, for the value $l/M \simeq 0.845$ (blue curve) the inner horizon and the event horizon of the black hole combine into a single horizon, i.e., one gets an extreme black hole. If the value of $l$ is even bigger the black hole horizons disappear and only the cosmological horizon remains. This is illustrated for $l/M =1$ (green curve). 
 The particular values used in this plot for the other parameters are $\{M/M_P = 100, a/M_P = 80, \Lambda_0 l_P^2 = 10^{-6}\}$.}
\end{figure}

On the other hand, only if quantum modifications were of the order of the mass of the black hole ($l\sim M$), a quantum improved black hole could have rather different inner and outer horizons. In this way, if a classical black hole possesses the three typical horizons, its quantum improved counterpart, provided with the same mass, rotation parameter and $\Lambda_0$, could have similar horizons or it could also be an extreme black hole (only a BH horizon and a cosmological horizon) or it could even not be a black hole (hyperextreme case, where only the cosmological horizon remains). This is illustrated in figure \ref{BigQuantumCH}.

%

Let us remark that the explicit profiles in (\ref{model}) are not unique; they are chosen as a minimal illustrative interpolation between the UV and IR limits. More general interpolating functions satisfying the regularity conditions of Theorem 1 and the correct asymptotic limits would still define scalar-regular rotating black holes in our class. Their main effect would be quantitative: they would modify the finite interior curvature scales and, through the zeros of $\Delta_r$, shift the location and even the number of horizons, especially in the transition region and near extremality. By contrast, for the minimal-order class with non-negative mass function, the qualitative conclusion on weak-energy-condition violation remains unchanged by Theorem 2.

\section{Conclusions}\label{Conclu}

Under the working assumption that an effective Lorentzian manifold endowed with a smooth metric provides an adequate description of a regular rotating black hole embedded in a dark-energy environment (i.e., a potentially scale-dependent $\Lambda$
background), we have constructed the most general stationary and axisymmetric geometry within the generalized Kerr--Schild class studied in this work and analyzed its main geometric and physical properties. 
Within this framework, the absence of scalar curvature singularities 
is equivalent to explicit constraints on the differentiability and near-core behaviour of the mass profile \(\mathcal{M}(r)\) and of the 
cosmological term \(\Lambda(r)\).
We further showed that, under the minimal regularity assumptions adopted here, enforcing scalar regularity generically entails violations of the standard pointwise energy conditions, indicating that effective non-classical stresses are required in the deep interior.
Finally, since scalar-regular rotating black holes in our class do not require an extension through the would-be disk, we consistently restrict to a topologically trivial spacetime; within this setting, causal pathologies can be excluded under simple sufficient conditions, namely \(\mathcal{M}(r)\ge 0\) and \(\Lambda(r)\ge -3/a^2\) throughout the domain of interest, which guarantees stable causality in the spacetime.

Since the generalized Newman--Janis algorithm is not directly applicable when a dark-energy sector (equivalently, a cosmological constant or a running \(\Lambda\)-term) is present, in Sec.\ref{SSTRBH} we developed an alternative and fully explicit construction that lifts regular, static, spherically symmetric black-hole models in a dark-energy background to regular rotating spacetimes within our generalized Kerr--Schild framework.
This procedure provides a systematic map from the input functions characterizing the seed spherically symmetric solution---in particular the mass profile and the \(\Lambda\)-sector---to the corresponding stationary, axisymmetric geometry, thereby enabling the generation and analysis of broad families of regular rotating black holes beyond the reach of Newman--Janis--type methods.

An essential caveat concerns dynamical stability: our analysis focuses on the stationary geometry and does not assess how perturbations may alter the maximal extension, in particular in the vicinity of the inner (Cauchy) horizon. Inner-horizon instability is well established for Reissner--Nordstr\"om (mass inflation) and Kerr spacetimes \cite{P&I,B&W,P&I2}. Most regular black-hole constructions with nonzero inner-horizon surface gravity $\kappa_-$ likewise retain an inner horizon that can be destabilized even by backscattered Hawking flux \cite{TorresIns,Cetal1,Cetal2,Cetal3}, with only finely tuned $\kappa_-=0$ exceptions \cite{Cetal3,FLMV}. 
However, even in that case semiclassical effects may still reintroduce an instability \cite{McMaken:2023ugd}. 

In the present framework, a running $\Lambda(r)$ modifies the metric functions that determine the location and surface gravity of the inner horizon. In our explicit model, $\Lambda(r)$ decreases monotonically from its UV value toward its IR value, so the $\Lambda'(r)$ contribution to $\kappa \propto \Delta'_r$ tends to oppose the classical negative slope at the inner horizon. This suggests that a running-$\Lambda(r)$ sector could soften the instability quantitatively by reducing the magnitude of $\kappa_{IH}$ in some regions of parameter space. However, the net effect is model-dependent. A systematic stability analysis across the proposed class of spacetimes would require a dedicated perturbative/backreaction analysis, which lies beyond the scope of the present work \cite{CarballoRubio:2024ztw}. Therefore, this constitutes a natural and important direction for future work.

Regarding the observational validation of the models presented here, these results provide a controlled stationary-axisymmetric framework for future numerical studies of geodesics, shadows and horizon-scale observables of regular rotating black holes in dark-energy backgrounds. We emphasize two key aspects:

First, the recent DESI BAO measurements, when combined with CMB and SNe datasets, provide concrete, statistically non-trivial indications that dark energy might be evolving, in qualitative agreement with the expectations of running-$\Lambda$ scenarios. While $\Lambda$CDM remains a viable benchmark model, ongoing and upcoming surveys (DESI extension, Euclid, the Nancy Grace Roman Space Telescope, Rubin/LSST, CMB-S4) will dramatically sharpen the picture. They will either reinforce the case for a running $\Lambda$ and other forms of DDE or push the allowed level of vacuum dynamics toward zero, providing an unprecedented test of the deepest interface between gravity, quantum theory, and the large-scale structure of the Universe.

Second, rapid progress in black-hole observations---from the LIGO--Virgo--KAGRA gravitational wave network to the Event Horizon Telescope, and soon LISA---is beginning to test theoretical predictions for rotating black holes embedded in dark-energy environments. At the current level of sensitivity and angular resolution, the data remain compatible with both the solutions of General Relativity and a broad class of alternative regular RBH scenarios. However, as measurement precision improves and multi-messenger constraints accumulate, these facilities offer a realistic prospect of discriminating among competing models and of turning the rotating RBH phenomenology discussed here into a quantitatively testable framework.


\newpage
\appendix

\section*{Appendix}

When the ring is approached with finite $\tilde\xi$ the Kretschmann scalar behaves as

\begin{equation*}
\resizebox{\linewidth}{!}{$
\begin{aligned}
\mathcal{K}
  &\rightarrow\frac{1}{%
        9\bigl(\Lambda(0)\,a^{2}+3\bigr)^{2}%
        \bigl(\tilde{\xi}^{2}+1\bigr)^{6}}%
  \Bigl(
     24\Lambda(0)^{2}\bigl(\Lambda(0)\,a^{2}+3\bigr)^{2}\,\tilde{\xi}^{12}
\\[2pt]
&\quad
  +24\Bigl(
       6\Lambda(0)^{2}a^{2}
      +\bigl(2\mathcal{M}'''(0)\,a^{2}+18\bigr)\Lambda(0)
      +\Lambda^{(4)}(0)\,a^{4}
      +6\mathcal{M}'''(0)
    \Bigr)
    \bigl(\Lambda(0)\,a^{2}+3\bigr)\Lambda(0)\,\tilde{\xi}^{10}
\\[2pt]
&\quad
  +\Bigl(
       360\Lambda(0)^{4}a^{4}
      +\bigl(240\mathcal{M}'''(0)\,a^{4}+2160a^{2}\bigr)\Lambda(0)^{3}
\\[-2pt] &\qquad
      +\bigl(120a^{6}\Lambda^{(4)}(0)
             +144\mathcal{M}'''(0)^{2}a^{4}
             +1440\mathcal{M}'''(0)\,a^{2}+3240\bigr)\Lambda(0)^{2}
\\[-2pt] &\qquad
      +\bigl(72\Lambda^{(4)}(0)\,a^{6}\mathcal{M}'''(0)
             +360\Lambda^{(4)}(0)\,a^{4}
             +864\mathcal{M}'''(0)^{2}a^{2}
             +2160\mathcal{M}'''(0)\bigr)\Lambda(0)
\\[-2pt] &\qquad
      +18\Lambda^{(4)}(0)^{2}a^{8}
      +216\Lambda^{(4)}(0)\mathcal{M}'''(0)\,a^{4}
      +1296\mathcal{M}'''(0)^{2}
    \Bigr)\tilde{\xi}^{8}
\\[2pt]
&\quad
  +\Bigl(
       480\Lambda(0)^{4}a^{4}
      +\bigl(480\mathcal{M}'''(0)\,a^{4}+2880a^{2}\bigr)\Lambda(0)^{3}
\\[-2pt] &\qquad
      +\bigl(240a^{6}\Lambda^{(4)}(0)
             -48\mathcal{M}'''(0)^{2}a^{4}
             +2880\mathcal{M}'''(0)\,a^{2}+4320\bigr)\Lambda(0)^{2}
\\[-2pt] &\qquad
      +\bigl(96\Lambda^{(4)}(0)\,a^{6}\mathcal{M}'''(0)
             +720\Lambda^{(4)}(0)\,a^{4}
             -288\mathcal{M}'''(0)^{2}a^{2}
             +4320\mathcal{M}'''(0)\bigr)\Lambda(0)
\\[-2pt] &\qquad
      +64\Lambda^{(4)}(0)^{2}a^{8}
      +288\Lambda^{(4)}(0)\mathcal{M}'''(0)\,a^{4}
      -432\mathcal{M}'''(0)^{2}
    \Bigr)\tilde{\xi}^{6}
\\[2pt]
&\quad
  +\Bigl(
       360\Lambda(0)^{4}a^{4}
      +\bigl(480\mathcal{M}'''(0)\,a^{4}+2160a^{2}\bigr)\Lambda(0)^{3}
\\[-2pt] &\qquad
      +\bigl(240a^{6}\Lambda^{(4)}(0)
             +264\mathcal{M}'''(0)^{2}a^{4}
             +2880\mathcal{M}'''(0)\,a^{2}+3240\bigr)\Lambda(0)^{2}
\\[-2pt] &\qquad
      +\bigl(96\Lambda^{(4)}(0)\,a^{6}\mathcal{M}'''(0)
             +720\Lambda^{(4)}(0)\,a^{4}
             +1584\mathcal{M}'''(0)^{2}a^{2}
             +4320\mathcal{M}'''(0)\bigr)\Lambda(0)
\\[-2pt] &\qquad
      +84\Lambda^{(4)}(0)^{2}a^{8}
      +288\Lambda^{(4)}(0)\mathcal{M}'''(0)\,a^{4}
      +2376\mathcal{M}'''(0)^{2}
    \Bigr)\tilde{\xi}^{4}
\\[2pt]
&\quad
  +\Bigl(
       144\Lambda(0)^{4}a^{4}
      +\bigl(240\mathcal{M}'''(0)\,a^{4}+864a^{2}\bigr)\Lambda(0)^{3}
\\[-2pt] &\qquad
      +\bigl(120a^{6}\Lambda^{(4)}(0)
             +96\mathcal{M}'''(0)^{2}a^{4}
             +1440\mathcal{M}'''(0)\,a^{2}+1296\bigr)\Lambda(0)^{2}
\\[-2pt] &\qquad
      +96\bigl(\Lambda^{(4)}(0)\,a^{4}+6\mathcal{M}'''(0)\bigr)
             \bigl(\mathcal{M}'''(0)\,a^{2}+\tfrac{15}{4}\bigr)\Lambda(0)
\\[-2pt] &\qquad
      +48\Lambda^{(4)}(0)^{2}a^{8}
      +288\Lambda^{(4)}(0)\mathcal{M}'''(0)\,a^{4}
      +864\mathcal{M}'''(0)^{2}
    \Bigr)\tilde{\xi}^{2}
\\[2pt]
&\quad
  +24\Lambda(0)^{4}a^{4}
  +\bigl(48\mathcal{M}'''(0)\,a^{4}+144a^{2}\bigr)\Lambda(0)^{3}
\\[-2pt] &\quad
  +\bigl(24a^{6}\Lambda^{(4)}(0)
         +24\mathcal{M}'''(0)^{2}a^{4}
         +288\mathcal{M}'''(0)\,a^{2}+216\bigr)\Lambda(0)^{2}
\\[-2pt] &\quad
  +24\bigl(\mathcal{M}'''(0)\,a^{2}+3\bigr)
        \bigl(\Lambda^{(4)}(0)\,a^{4}+6\mathcal{M}'''(0)\bigr)\Lambda(0)
\\[-2pt] &\quad
  +10\Lambda^{(4)}(0)^{2}a^{8}
  +72\Lambda^{(4)}(0)\mathcal{M}'''(0)\,a^{4}
  +216\mathcal{M}'''(0)^{2}
  \Bigr)
\end{aligned}
$}
\end{equation*}

and when the ring is approached with infinite $\tilde\xi$ it behaves as $\mathcal K \rightarrow (8/3) \Lambda(0)^2$. 

When the ring is approached with finite $\tilde\xi$ the invariant $w_1$ behaves as 

\begin{equation*}
\begin{alignedat}{1}
w_{1}\;\rightarrow\;
&\frac{1}{%
  24\bigl(\Lambda(0)a^{2}+3\bigr)^{2}(\tilde{\xi}^{2}+1)^{6}}
\Bigl(\\[2pt]
&(\Lambda^{(4)}(0)^{2}a^{8}
  +4\mathcal M'''(0)a^{6}\Lambda^{(4)}(0)\Lambda(0)\\
&\quad
  +(16\mathcal M'''(0)^{2}\Lambda(0)^{2}
    +12\mathcal M'''(0)\Lambda^{(4)}(0))a^{4}\\
&\quad
  +96\Lambda(0)\mathcal M'''(0)^{2}a^{2}
  +144\mathcal M'''(0)^{2})\,\tilde{\xi}^{8}\\[4pt]
&-16\mathrm{I}\,
  (\Lambda(0)a^{2}+3)\mathcal M'''(0)\,
  (\Lambda^{(4)}(0)a^{4}
   +4\mathcal M'''(0)\Lambda(0)a^{2}
   +12\mathcal M'''(0))\,\tilde{\xi}^{7}\\[4pt]
&+4(\Lambda^{(4)}(0)a^{4}
     +3\mathcal M'''(0)\Lambda(0)a^{2}
     +9\mathcal M'''(0))\\
&\quad\;\,
  (\Lambda^{(4)}(0)a^{4}
   -8\mathcal M'''(0)\Lambda(0)a^{2}
   -24\mathcal M'''(0))\,\tilde{\xi}^{6}\\[4pt]
&+64\mathrm{I}\,
  (\Lambda(0)a^{2}+3)^{2}\mathcal M'''(0)^{2}\,\tilde{\xi}^{5}\\[4pt]
&+6(\Lambda^{(4)}(0)a^{4}
     -2\mathcal M'''(0)\Lambda(0)a^{2}
     -6\mathcal M'''(0))\\
&\quad\;\,
  (\Lambda^{(4)}(0)a^{4}
   -\tfrac43\mathcal M'''(0)\Lambda(0)a^{2}
   -4\mathcal M'''(0))\,\tilde{\xi}^{4}\\[4pt]
&+16\mathrm{I}\,
  a^{4}(\Lambda(0)a^{2}+3)\mathcal M'''(0)\Lambda^{(4)}(0)\,\tilde{\xi}^{3}\\[4pt]
&+4a^{4}\Lambda^{(4)}(0)\,
  (\Lambda^{(4)}(0)a^{4}
   +\mathcal M'''(0)\Lambda(0)a^{2}
   +3\mathcal M'''(0))\,\tilde{\xi}^{2}\\[4pt]
&+\Lambda^{(4)}(0)^{2}a^{8}
\Bigr)
\end{alignedat}
\end{equation*}

while for infinite  $\tilde{\xi}$ it just satisfies $w_1\rightarrow 0$.

(In the absence of $\Lambda$, one gets, as a subcase, the known results \cite{TorresReg}
\[
w_1\rightarrow \frac{2 \tilde{\xi}^4 \mathcal M'''(0)^2}{3 (1-I \tilde{\xi})^4 (1+\tilde{\xi}^2)^2}
\]
for finite $\tilde{\xi}$ and $w_1\rightarrow 0$ for infinite $\tilde{\xi}$).



\end{document}